% hep-th v1
\input harvmac
%\input epsf
%\draftmode

\overfullrule=0pt
\abovedisplayskip=12pt plus 3pt minus 1pt
\belowdisplayskip=12pt plus 3pt minus 1pt
%macros
%
\def\tilde{\widetilde}

\def\to{\rightarrow}
\def\tphi{{\tilde\phi}}

\def\bigone{\hbox{1\kern -.23em {\rm l}}}
\def\ZZ{\hbox{\zfont Z\kern-.4emZ}}
\def\half{{\litfont {1 \over 2}}}

\font\litfont=cmr6

\def\tX{{\tilde X}}

\def\tpa{{2\pi\alpha'}}

\def\ola{\overleftarrow}
\def\ora{\overrightarrow}
\def\tPsi{{\tilde\Psi}}

\lref\seiwit{N. Seiberg and E. Witten, {\it ``String Theory and
Noncommutative Geometry''}, hep-th/9908142, JHEP {\bf 9909}, 032 (1999).}
\lref\seibnew{N. Seiberg, {\it ``A Note on Background
Independence in Noncommutative Gauge Theories, Matrix Model and
Tachyon Condensation''}, hep-th/0008013, JHEP {\bf 0009}, 003 (2000).}
%\lref\smnvs{S. Mukhi and N.V. Suryanarayana, {\it ``Chern-Simons 
%Terms on Noncommutative Branes''}, hep-th/0009101, 
%JHEP {\bf 0011}, 006 (2000).}
\lref\garousi{M. Garousi, {\it ``Noncommutative World Volume
Interactions on D-branes and Dirac-Born-Infeld Action''},
hep-th/9909214, Nucl. Phys. {\bf B579} 209 (2000).}
\lref\wyllard{N. Wyllard, {\it ``Derivative Corrections to D-brane 
Actions with Constant Background  Fields''}, hep-th/0008125,
Nucl. Phys. {\bf B598}, 247 (2001).}
\lref\earlyderivcomp{
J.H.~Schwarz, {\it ``Superstring Theory''}, Physics Reports {\bf 89},
223 (1982)\semi A.A. Tseytlin, {\it ``Vector Field Effective Actions
in the Open Superstring Theory''}, Nucl. Phys. {\bf B276}, 391
(1986)\semi 
O.D. Andreev and A.A. Tseytlin, {\it ``Partition Function
Representation for the Open Superstring Effective Action: Cancellation
of M\"obius Infinities and Derivative Corrections to the Born-Infeld
Lagrangian''}, Nucl. Phys. {\bf B311}, 205 (1988)\semi
K.~Hashimoto, {\it ``Corrections to D-brane action and generalized
boundary state''}, Phys. Rev. {\bf D61} 106002 (2000),
hep-th/9909027\semi
K. Hashimoto, {\it ``Generalized Supersymmetric Boundary State''},
hep-th/9909095, JHEP {\bf 0004}, 023 (2000).}
\lref\dms{S.R. Das, S. Mukhi and N.V. Suryanarayana, {\it ``Derivative 
Corrections from Noncommutativity''}, hep-th/0106024.} 
\lref\callan{C.~G. Callan, C.~Lovelace, C.~R.~Nappi 
and S.~A. Yost, {\it ``String loop   corrections to beta functions''},
Nucl. Phys. {\bf B288} 525 (1987)\semi
C.~G. Callan, C.~Lovelace, C.~R.~Nappi, and S.~A. Yost, 
{\it ``Adding holes and   crosscaps to the superstring''},
Nucl.~Phys.~{\bf B293} 83 (1987)\semi
C.~G. Callan, C.~Lovelace, C.~R. Nappi, and S.~A. Yost, 
{\it ``Loop corrections to   superstring equations of motion''}
Nucl. Phys. {\bf B308} (1988) 221.}
\lref\divecchia{P.~{Di Vecchia} and A.~Liccardo, 
{\it ``D branes in string theory, I''}, hep-th/9912161\semi
P.~{Di Vecchia} and A.~Liccardo, 
{\it ``D branes in string theory, II''}, hep-th/9912275.}

{\nopagenumbers
\Title{\vbox{
\hbox{hep-th/0108072}
\hbox{TIFR/TH/01-28}}}
{\centerline{Star Products from Commutative String Theory}}
\centerline{Sunil Mukhi}
\vskip 8pt
\centerline{\it Tata Institute of Fundamental Research,}
\centerline{\it Homi Bhabha Rd, Mumbai 400 005, India}

\vskip 2truecm
\centerline{\bf ABSTRACT}
\medskip
A boundary-state computation is performed to obtain derivative
corrections to the Chern-Simons coupling between a $p$-brane and the
RR gauge potential $C^{p-3}$. We work to quadratic order in the gauge
field strength $F$, but all orders in derivatives. In a certain limit,
which requires the presence of a constant $B$-field background, it is
found that these corrections neatly sum up into the $*_2$ product of
(commutative) gauge fields. The result is in agreement with a recent
prediction using noncommutativity.

\vfill
\Date{August 2001}
\eject}
\ftno=0

%\listtoc
%\writetoc

\noindent{\bf Introduction}
\medskip

In a recent paper\refs\dms\ it was shown that the noncommutative
formulation of open-string theory can actually give detailed information
about ordinary commutative string theory. Once open Wilson lines are
included in the noncommutative action, one has exact equality of
commutative and noncommutative actions including all $\alpha'$
corrections on both sides. As a result, a lot of information about
$\alpha'$ corrections on the commutative side is encoded in the
lowest-order term (Chern-Simons or DBI) on the noncommutative side,
and can be extracted explicitly.

The predictions of Ref.\refs\dms\ were tested against several
boundary-state computations in commutative open-string theory
performed in Ref.\refs\wyllard, and impressive agreement was
found. The latter calculations were restricted to low-derivative
orders, largely because the boundary-state computation becomes rather
tedious when we go to high derivative order. However, in some specific
cases, particularly when focusing on Chern-Simons couplings in the
Seiberg-Witten limit\refs\seiwit, the predictions from
noncommutativity in Ref.\refs\dms\ are simple and elegant to all
derivative orders as long as we work with weak field strengths
(quadratic order in $F$). This suggests that the boundary state
computation can be performed for these special cases, and in the given
limits, to all derivative orders.

In this short note, we perform precisely such a calculation, using
techniques and formulae already established in Ref.\refs\wyllard. It
will turn out that the derivative corrections neatly sum up and give
rise to a $*_2$ product\refs{\garousi} between a pair of commutative
field strengths:
\eqn\startwodef{
\langle F_{ij}(x),F_{kl}(x) \rangle_{*_2} \equiv 
F_{ij}(x){\sin(\half\ola{~\partial_p}\,\theta^{pq}\ora{\,\partial_q\,})
\over \half\ola{~\partial_p}\,\theta^{pq}\ora{\,\partial_q\,} } 
F_{kl}(x) }
The expression obtained in this way for the derivative
corrections agrees perfectly with a prediction from noncommutativity
that was made in Ref.\refs\dms. 

Besides verifying this prediction, the calculation described here
suggests that derivative corrections to brane actions in commutative
string theory, even away from the Seiberg-Witten limit, might have a
novel underlying mathematical structure. We will comment on this at the
end.
\bigskip

\noindent{\bf Chern-Simons Corrections: RR 6-form}
\medskip

In this section, we compute the corrections to the term 
\eqn\cssix{
S_{CS} = \half\int C_{RR}^{(6)}\wedge F\wedge F }
on a Euclidean D9-brane of type IIB string theory with
noncommutativity along all 10 directions. Here $C^{6}$ is the
Ramond-Ramond 6-form potential.  The computation is performed to all
orders in the derivative expansion, but keeping only terms of order
$(F^2)$. The use of D9-branes is purely a convenience, the same
calculation can be trivially applied to D$p$-branes and their coupling
to the RR form $C^{(p-3)}$.

The computation of corrections will be done in the boundary-state
formalism. Useful background on how to compute derivative corrections
in this formalism may be found in Ref.\refs\wyllard. The formalism
itself was developed in Ref.\refs\callan, and has been reviewed
recently in Ref.\refs\divecchia. Earlier work on derivative
corrections can be found in Refs.\refs\earlyderivcomp.

Let us denote the sum of all derivative corrections to $S_{CS}$ as
$\Delta S_{CS}$. Our starting point is the expression
\eqn\wylstart{
S_{CS} + \Delta S_{CS} =
\big\langle C \big| e^{-{i\over \tpa} \int d\sigma d\theta D\phi^\mu
A_\mu(\phi) } \big|B\big\rangle_{R} }
where $|C\rangle$ represents the RR field, and $|B\rangle_R$ is the
Ramond-sector boundary state for zero field strength. We are using
superspace notation, for example $\phi^\mu = X^\mu + \theta \psi^\mu$
and $D$ is the supercovariant derivative.

Combining Eqs.(2.3),(2.6),(2.13) of Ref.\refs\wyllard, we can
rewrite this as:
\eqn\wylnext{
\eqalign{
S_{CS} + \Delta S_{CS} &=
\big\langle C\big| 
e^{{i\over \tpa} \int d\sigma d\theta\sum_{k=0}^\infty{1\over (k+1)!}
{k+1\over k+2} D\tphi^\nu\tphi^\mu \tphi^{\lambda_1}\cdots
\tphi^{\lambda_k}\del_{\lambda_1}\ldots\del_{\lambda_k}
F_{\mu\nu}(x)}\times\cr
&\phantom{\langle C|}
e^{{i\over\tpa} \int d\sigma[\tPsi^\mu \psi_0^\nu
+ \psi_0^\mu \psi_0^\nu]\sum_{k=0}^\infty{1\over k!}\tX^{\lambda_1}
\cdots \tX^{\lambda_k} \del_{\lambda_1}\ldots\del_{\lambda_k} 
F_{\mu\nu}(x)}\big|B\big\rangle_{R}\cr} }
where nonzero modes have a tilde on them, while the zero modes are
explicitly indicated.

Since we are looking for couplings to the RR 6-form $C^{(6)}$, and
working to order $F^2$, we only need terms with the structure
$\del\ldots\del F\wedge\del\ldots\del F$. For such terms, two $F$'s
and 4 $\psi_0$'s must be retained. Thus we can drop the first
exponential factor in Eq.\wylnext\ above, as well as the first fermion
bilinear $\tPsi^\mu \psi_0^\nu$ in the second exponential. Then,
expanding the exponential to second order, we get:
\eqn\exposecond{
\eqalign{
S_{CS} + \Delta S_{CS} =& ~
\half\sum_{n=0}^\infty \sum_{p=0}^\infty \left({i\over\tpa}\right)^2
\int_0^{2\pi} \!\! d\sigma_1\int_0^{2\pi} \!\! d\sigma_2 ~
\big\langle C\big| \left(\half\psi_0^\mu\psi_0^\nu\right)
\left(\half\psi_0^\alpha\psi_0^\beta\right)~\times\cr
& {1\over n!} \tX^{\lambda_1}(\sigma_1) \cdots \tX^{\lambda_n}(\sigma_1)
{1\over p!} \tX^{\rho_1}(\sigma_2) \cdots \tX^{\rho_p}(\sigma_2)~\times\cr
&\del_{\lambda_1}\ldots\del_{\lambda_n} F_{\mu\nu}(x)\,
\del_{\rho_1}\ldots\del_{\rho_p} F_{\alpha\beta}(x)
\big|B\big\rangle_{R} \cr}}
Now we need to evaluate the 2-point functions of the $\tX$. The
relevant contributions have non-logarithmic finite parts\refs\wyllard\
and come from propagators for which there is no self-contraction. This
requires that $n=p$. Then we get a combinatorial factor of $n!$ from
the number of such contractions in $\big\langle
\big(\tX(\sigma_1)\big)^n\,
\big(\tX(\sigma_2)\big)^n\big\rangle$. The result is:
\eqn\contrac{
\eqalign{
S_{CS} + \Delta S_{CS} =& ~
 \half
\sum_{n=0}^\infty
{1\over n!} \left({i\over\tpa}\right)^2
\int_0^{2\pi} \!\! d\sigma_1\int_0^{2\pi} \!\! d\sigma_2 ~
D^{\lambda_1\rho_1}(\sigma_1-\sigma_2)\cdots 
D^{\lambda_n\rho_n}(\sigma_1-\sigma_2)~\times\cr
&\del_{\lambda_1}\ldots\del_{\lambda_n} F_{\mu\nu}(x)\,
\del_{\rho_1}\ldots\del_{\rho_n} F_{\alpha\beta}(x)
~\big\langle C\big|\left(\half\psi_0^\mu\psi_0^\nu\right)
\left(\half\psi_0^\alpha\psi_0^\beta\right)\big|B\big\rangle_{R}\cr }}
The fermion zero mode expectation values are evaluated using the
recipe:
\eqn\zerorecipe{
\half \psi_0^\mu\psi_0^\nu F_{\mu\nu} \rightarrow
(-i\alpha')F }
where the $F$ on the right hand side is a differential 2-form.
The justification for this can be found below Eq.(B.3) of 
Ref.\refs\wyllard. Thus we are led to:
\eqn\fermizero{
S_{CS} + \Delta S_{CS} =
T^{\lambda_1\ldots\lambda_n;\,\rho_1\ldots\rho_n} 
\,\del_{\lambda_1}\ldots\del_{\lambda_n} F \wedge
\del_{\rho_1}\ldots\del_{\rho_n} F}
where
\eqn\ttensor{
T^{\lambda_1\ldots\lambda_n;\,\rho_1\ldots\rho_n} \equiv~ 
\half {1\over n!}\left({i\over\tpa}\right)^2 (-i\alpha')^2  
\int_0^{2\pi} \!\! d\sigma_1\int_0^{2\pi} \!\! d\sigma_2 \,
D^{\lambda_1\rho_1}(\sigma_1-\sigma_2)\cdots 
D^{\lambda_n\rho_n}(\sigma_1-\sigma_2)}

Now we insert the expression for the propagator:
\eqn\propagat{
D^{\mu\nu}(\sigma_1 - \sigma_2)
= \alpha'\sum_{m=1}^\infty {e^{-\epsilon m}\over m}
\left(h^{\mu\nu}e^{im(\sigma_2-\sigma_1)}
+ h^{\nu\mu}e^{-im(\sigma_2-\sigma_1)} \right)}
where $\epsilon$ is a regulator, and
\eqn\hmunu{
h^{\mu\nu} \equiv {1\over g + \tpa(B+F)} }
As is well known, the propagator is no longer symmetric when a
$B$-field background is turned on. We now find that:
\eqn\evalu{
\eqalign{
T^{\lambda_1\ldots\lambda_n;\,\rho_1\ldots\rho_n}
=~&\half {1\over n!} (\alpha')^n 
\sum_{m_1=1}^\infty \cdots \sum_{m_n=1}^\infty ~
{e^{-\epsilon (m_1+\cdots m_n)}\over m_1\ldots m_n}~\times\cr
&
\int_0^{2\pi}{d\sigma_1\over 2\pi}
\int_0^{2\pi} {d\sigma_2\over 2\pi}~
\prod_{i=1}^n \left(h^{\lambda_i\rho_i}e^{im_i(\sigma_2-\sigma_1)}
+ h^{\rho_i\lambda_i}e^{-im_i(\sigma_2-\sigma_1)} \right) \cr}}
It is convenient to define 
$$
(h^+)^{\mu\nu} \equiv h^{\mu\nu},\qquad 
(h^-)^{\mu\nu} \equiv h^{\nu\mu} 
$$
which allows us to write:
$$
\left(h^{\mu\nu}e^{im(\sigma_2-\sigma_1)}
+ h^{\nu\mu}e^{-im(\sigma_2-\sigma_1)} \right)
= \sum_\pm (h^\pm)^{\mu\nu} e^{\pm im(\sigma_2-\sigma_1)} 
$$
and we find that
\eqn\evalutwo{
T^{\lambda_1\ldots\lambda_n;\,\rho_1\ldots\rho_n}
=\half {1\over n!} (\alpha')^n 
\int_0^{2\pi} {d\sigma\over 2\pi}~
\prod_{i=1}^n ~\Big(
\sum_\pm (h^\pm)^{\lambda_i\rho_i} \sum_{m=1}^\infty 
{e^{-\epsilon m}\over m}e^{\pm im\sigma}\Big)}
After evaluating the sum over $m$, the result, depending on the
regulator $\epsilon$, is
\eqn\evalusummed{
T^{\lambda_1\ldots\lambda_n;\,\rho_1\ldots\rho_n}
=\half {1\over n!} (\alpha')^n
\int_0^{2\pi} {d\sigma\over 2\pi}~
\prod_{i=1}^n ~\left(
-\sum_\pm (h^\pm)^{\lambda_i\rho_i} 
\ln (1-e^{-\epsilon\pm i\sigma})\right)}

At this point it proves difficult to proceed further without 
introducing some simplification. The integral above, for general
$h^{\mu\nu}$, can only be performed explicitly for $n=2$, as has in
fact been done in Ref.\refs\wyllard. However, if we take a limit where
\eqn\swlimit{
g_{\mu\nu} \sim \delta,\qquad B_{\mu\nu}\sim {\rm fixed},
\qquad \alpha'\sim \sqrt\delta}
with $\delta\rightarrow 0$, a simplification occurs. This is indeed
just the Seiberg-Witten limit\refs\seiwit. In this limit, the
``metric'' $h_{\mu\nu}$ becomes antisymmetric:
\eqn\metriclim{
h^{\mu\nu}\to{\theta^{\mu\nu}\over \tpa} }
where
\eqn\thetadef{
\theta^{\mu\nu}\equiv \left({1\over B}\right)^{\mu\nu} }
and hence we find:
\eqn\swlim{
\eqalign{
\sum_\pm (h^\pm)^{\lambda_i\rho_i} \ln (1-e^{-\epsilon\pm i\sigma})
&= {1\over\tpa} \theta^{\lambda_i\rho_i}\ln\left(
{1-e^{-\epsilon+ i\sigma}\over 1-e^{-\epsilon- i\sigma}}\right)\cr
&= {1\over\tpa}\, i\,(\sigma-\pi)\,\theta^{\lambda_i\rho_i}}}
The integrand has simplified considerably and the integral can now be
done. Also, we have now taken the regulator $\epsilon$ to 0, as it is
no longer needed. It follows that:
\eqn\ttensorsw{
\eqalign{
T^{\lambda_1\ldots\lambda_n;\,\rho_1\ldots\rho_n}
&=\half {1\over n!} \left(-{i\over 2\pi}\right)^n
\theta^{\lambda_1\rho_1}\ldots \theta^{\lambda_n\rho_n}
\int_0^{2\pi} {d\sigma\over 2\pi} (\sigma-\pi)^n\cr
&=\cases{ \half {1\over n!} \left(-{i\over 2\pi}\right)^n
{\pi^n\over n+1}~
\theta^{\lambda_1\rho_1}\ldots \theta^{\lambda_n\rho_n}
\quad({\rm even}~n)\cr\cr
 0\quad({\rm odd}~n)} }}

Inserting this back in Eq.\fermizero, it follows that keeping all
derivative orders, but restricting to quadratic order in $F$, and in
the Seiberg-Witten limit,
\eqn\derivsw{
\eqalign{
S_{CS} &+ \Delta S_{CS}\cr &= \half\int C^{(6)}\wedge \sum_{j=0}^\infty
(-1)^j {1\over 2^{2j} (2j+1)!} \,
\theta^{\lambda_1\rho_1}\ldots \theta^{\lambda_{2j}\rho_{2j}}\,
\del_{\lambda_1}\ldots\del_{\lambda_{2j}} F \wedge
\del_{\rho_1}\ldots\del_{\rho_{2j}} F \cr
&= 
\half\int C^{(6)}\wedge \langle F\wedge F\rangle_{*_2} }}
where the product $*_2$ was defined in Eq.\startwodef.

This agrees with a prediction from noncommutativity made in
Ref.\refs\dms, see Eq.(4.13) of that paper. In that sense, the result
is not surprising. However, it is amusing that using the
boundary-state formalism in ordinary (commutative) string theory, we
were explicitly able to obtain the $*_2$ product without invoking
noncommutativity in any form.
\bigskip

\noindent{\bf Conclusions}
\medskip

It should be reasonably straightforward to repeat the calculation
above to compute derivative corrections to 
$\int C^{(10-2n)}\wedge (F)^n$ for $n=3,4,5$ 
restricting to corrections of order $F^n$.  In
the Seiberg-Witten limit, one should find the $*_n$ product in this
way for these values of $n$. The analogous calculation for the DBI
action will perhaps be more difficult.

One of the most interesting questions raised by this calculation and
the work in Ref.\refs\dms\ is, what is the full expression for the
derivative corrections, away from the Seiberg-Witten limit. We know
that general string amplitudes depend on transcendental numbers, for
example $\zeta$-functions of odd argument. As noted in Ref.\refs\dms,
the Seiberg-Witten limit causes these to go away in all the cases
examined, leading to much simpler results which can then be recovered
using noncommutativity or, as in the present note, explicit boundary
state calculation. Clearly these simpler results place a strong
constraint on the form of the full derivative corrections, away from
the Seiberg-Witten limit. The question is then whether this constraint
can be combined with other inputs, such as boundary-state
computations, gauge invariance and
background-independence\refs{\seiwit,\seibnew}, to recover the full
corrections.  This could have important consequences in understanding
string theory beyond the derivative expansion.
\bigskip

\noindent{\bf Acknolwedgements}
\medskip

I am grateful to Sumit Das, Rajesh Gopakumar, Gautam Mandal, Shiraz
Minwalla, Ashoke Sen and Nemani Suryanarayana for helpful discussions.

\listrefs
\end